# Magnonics: Spin Waves Connecting Charges, Spins and Photons

A. V. Chumak[1] and H. Schultheiss[2]

[1]Fachbereich Physik and Landesforschungszentrum OPTIMAS, Technische Universität Kaiserslautern, 67663 Kaiserslautern, Germany
[2]Institut für Ionenstrahlphysik und Materialforschung, Helmholtz-Zentrum Dresden-Rossendorf, D-01328 Dresden, Germany

E-mails:     chumak@physik.uni-kl.de,
             h.schultheiss@hzdr.de

Spin waves (SW) are the excitation of the spin system in a ferromagnetic condensed matter body. They are collective excitations of the electron system and, from a quasi-classical point of view, can be understood as a coherent precession of the electrons' spins. Analogous to photons, they are also referred to as magnons indicating their quasi-particle character. The collective nature of SWs is established by the short-range exchange interaction as well as the non-local magnetic dipolar interaction, resulting in coherence of SWs from mesoscopic to even macroscopic length scales. As one consequence of this collective interaction, SWs are "charge current free" and, therefore, less subject to dissipation caused by scattering with impurities on the atomic level. This is a clear advantage over diffusive transport in spintronics that not only uses the charge of an electron but also its spin degree of freedom. Any (spin) current naturally involves motion and, thus, scattering of electrons leading to excessive heating as well as losses. This renders SWs a promising alternative to electric (spin) currents for the transport of spin information - one of the grand challenges of condensed matter physics.

The possibilities of SWs being a new means of information carriers stimulated the emerging research field called "magnonics". Its name - inspired by the terms "spintronics" and "photonics" - indicates the main driving force behind magnonics: exploiting SWs for information processing to supersede electronics. For frequencies ranging from gigahertz to terahertz, the SW wavelengths can be as small as only a few nanometers, orders of magnitude smaller compared to electromagnetic waves. In addition, their transport properties exhibit a strong dependence on the magnetization configuration that can be non-volatile and still reconfigurable on a sub-nanosecond timescale.

The field of modern magnonics is very wide and covers a variety of emerging physical phenomena. Some of the magnonics research directions are briefly summarized in Fig. 1. We present the directions that are already well established as well as the directions that are just at their initial state of their development but show large potential. In addition, one can see in the figure that the material science, the development of the methodology as well as the investigations of the spin-wave physical phenomena form the basis for all the research directions.

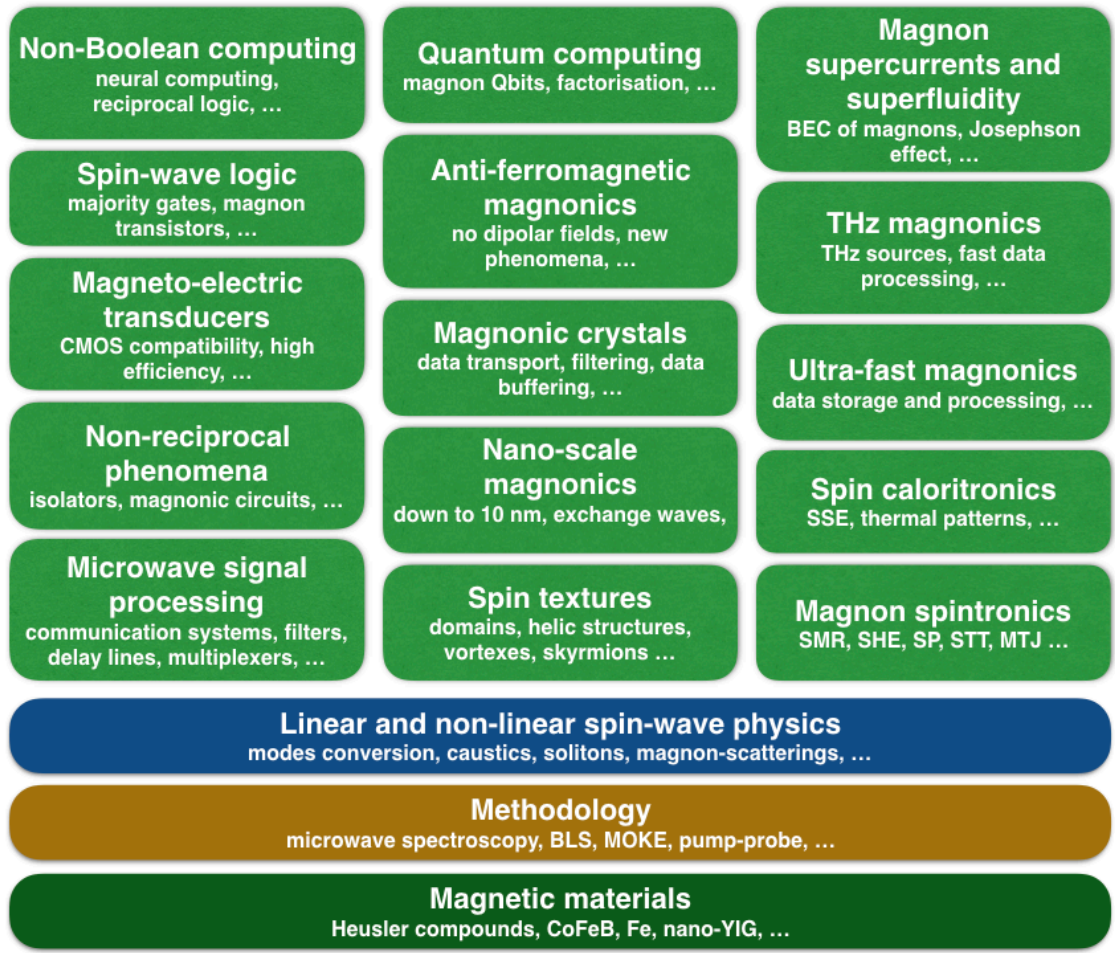

*Fig. 1. The variety of problems in modern magnonics.*

Below we discuss briefly the articles that are presented in this special issue on magnonics. They are associated with few different directions shown in Fig. 1. In particular with the "Magnetic materials", "Magnonic crystals", "Non-reciprocal phenomena", "Magnon spintronics", "Spin caloritronics", "Spin textures", and "Magnon supercurrents and superfluidity".

Spin-wave propagation requires utilization of magnetic materials with small magnetic loss, high values of saturation magnetization, and high Curie temperatures. Therefore, the search for new magnetic materials suitable for magnonics as well as improvement of existing materials is a constant process in the field. The growth of Yttrium Iron Garnet (YIG) films in nanometer thickness range become possible in the recent years and made a large impact on the field. The Liquid Phase Epitaxy (LPE) technique for the growth of YIG films with thicknesses of around 100 nm is presented in [1]. The surface roughness of the films is as low as 0.3 nm and a weak in-plane magnetic anisotropy with a pronounced six-fold symmetry is observed. The narrowest ferromagnetic resonance linewidth is determined to be 1.4 Oe @ 6.5 GHz and the Gilbert damping coefficient is estimated to be close to $1 \times 10^{-4}$.

One of the important part of the field of magnonics occupies investigations of magnonic crystals – artificial magnetic materials with properties periodically varied in space. Spectra of spin-wave excitations in such structures are significantly different compared to uniform media and exhibit features such as band gaps, where spin waves are not allowed to propagate. A spin-wave band spectrum of magnonic crystals formed by stacking thin film magnetic layers, with general assumptions about the properties of the interfaces between the layers, has been studied theoretically in [2]. It was found that the band gaps are a ubiquitous attribute of a weakened interlayer coupling and a finite interface anisotropy. The band gaps in such systems represent a legacy of the discrete spin-wave spectrum of the individual magnetic layers periodically stacked to form the magnonic crystal. Surface spin wave states in 1D planar crystals localized on the surfaces resulting from the breaking of the periodic structure have been investigated in [3]. Geometries with periodic changes of the anisotropy field and the magnonic crystal composed of Fe and Ni stripes have been considered. It was shown that a system with modulated anisotropy in an exchange regime is a direct analog of the electronic crystal, while for the surface states existing in bi-component magnonic crystals in dipolarexchange regime the spin waves show distinct differences in comparison to the electronic crystals. It was found that tuning of the strength of magnetization pinning, resulting from the surface anisotropy or dipolar effect, is vitally important for the existence of surface states in magnonic crystals. Finally, the third paper on magnonic crystal presented in this special issue is a review of the results obtained by the authors and is concentrated on the application of magnonic crystals for data processing and information technologies [4]. Different approaches for the realization of static, reconfigurable, and dynamic magnonic crystals are presented along with a variety of novel wave phenomena discovered in these crystals. Magnonic crystals constitute one of the key elements since they open access to novel multi-functional magnonic devices and can be used as filters, sensors, frequency- and time- inverters, data buffering elements, or components of logic gates.

Non-reciprocal spin-wave waveguides, in which spin waves propagating in opposite directions have different properties, are of high importance for microwave signal processing as well as for the solution of the problem of reflections in future magnonic circuits. An interesting approach of current-induced modulation of backward spin-waves in Permalloy microstructures was studies experimentally and using numerical simulations in [5]. Even with the smaller current injection of $5 \times 10^{10}$ A m$^{-2}$ into ferromagnetic microwires, the backward volume spin waves exhibit a gigantic 200 MHz frequency shift. The sign of the frequency shift is defined by the electric current direction and therefore opens access to the nonreciprocal magnonic conduits. The effective coupling between electric current and backward volume spin waves has a potential to build up a logic control method that encodes signals into the phase and amplitude of spin-waves. Another non-reciprocal spin-wave propagation associated with the nature of Daemon-Eshbach spin-wave mode was studies in [6]. The unidirectional spin wave heat conveyer effect has been investigated in a 200 nm thin yttrium iron garnet film using lock-in thermography. Comparing the results for the Damon–Eshbach and the backward volume modes the authors have shown that the origin of the asymmetric heat profile are indeed the non-reciprocal spin waves.

A new concept in information processing is technology that not solely rely on the electrons charge, but also control the electrons spin. This approach is of high interest in magnonics and is known as magnon spintronics. A study of the tunnel magneto-Seebeck effect in MgO based magnetic tunnel junctions is presented in [7]. The electrodes consist of CoFeB with in-plane magnetic anisotropy. The temperature gradients, which generate a voltage across the tunnel junctions layer stack are created using laser heating. The systematic analysis has shown, that the distribution of the temperature gradient is essential, to achieve high voltage signals and reasonable resulting tunnel magneto-Seebeck ratios. Furthermore, artefacts on the edges open up further possibilities of more complex heating scenarios for spincaloritronics in spintronic devices. The spin-Hall magnetoresistance (SMR) phenomena is studies in [8] in a multidomain helical spiral magnet $Cu_2OSeO_3$|Pt heterostructures. The SMR response the structure at 5 K, when the magnetic domains are almost frozen, has been compared to that at elevated temperatures, when domain walls move easily. At 5 K the SMR amplitude vanishes at low applied magnetic fields, while at 50 K it does not. This phenomenon is explained by the effect of the magnetic field on the domain structure of $Cu_2OSeO_3$. Recently spin structures such as spin helices and cycloids have attracted a lot of interest and were initiated by the discovery of the skyrmion lattice phase in non-centrosymmetric helical magnets. Finally, the review [9] addresses questions how spin helices and skyrmion lattices enrich the microwave characteristics of magnetic materials. When discussing perspectives for microwave electronics and magnonics the authors focus particularly on insulating materials as they avoid eddy current losses, offer low spin-wave damping, and might allow for electric field control of collective spin excitations. These studies are of interest in the view of low energy consumption magnonics.

Recent discoveries of spin superfluidity and room-temperature magnon supercurrents are of large interest in the light of fundamental research. A summary of the recent theoretical ideas on the Bose–Einstein condensation of magnons and possible superfluidity in yttrium iron garnet is presented in [10]. The review contains also a new development of these ideas and a brief description of experimental facts in section. The reviewed theory that explains some of the experimental observations predicts that the reflection symmetry of the magnon gas is spontaneously violated at Bose–Einstein condensation in thick films. In thin films the condensate is symmetric at low magnetic field and transits to the non-symmetric state at higher field. Dipolar interaction energy depends on the phase of the condensate wave function. In quasi-equilibrium it traps the phase. The review here our past works on magnetization transport in insulating magnets and also add new insights, with a particular focus on magnon transport. The Nambu–Goldstone bosons in insulating magnets, which are called magnons or spin waves and play a key role in magnetization transport, are discussed in [11]. The authors summarize in detail the magnon counterparts of electron transport, such as the Wiedemann–Franz law, the Onsager reciprocal relation between the Seebeck and Peltier coefficients, the Hall effects, the superconducting state, the Josephson effects, and the persistent quantized current in a ring to list a few. Focusing on the electromagnetism of moving magnons, i.e. magnetic dipoles, a way to directly measure magnon currents was proposed theoretically.

# References

[1] Dubs C, Surzhenko O, Linke R, Danilewsky A, Brückner U and Delith J 2017 *J. Phys. D: Appl. Phys.* **50** 204005.

[2] Kruglyak V V, Davies C S, Tkachenko V S, Gorobets O Yu, Gorobets Yu I and Kuchko A N 2017 *J. Phys. D: Appl. Phys.* **50** 094003.

[3] Rychly J and Klos J W 2017 *J. Phys. D: Appl. Phys.* **50** 164004.

[4] Chumak A V, Serga A A and Hillebrands B 2017 *J. Phys. D: Appl. Phys.* **50** 244001.

[5] Sato N, Lee S-W, Lee K-J and Sekiguchi K 2017 *J. Phys. D: Appl. Phys.* **50** 094004.

[6] Wid O, Bauer J, Müller A, Breitenstein O, Parkin S S P and Schmidt G 2017 *J. Phys. D: Appl. Phys.* **50** 134001.

[7] Martens U, Walowski J, Schumann T, Mansurova M, Boehnke A, Huebner T, Reiss G, Thomas A and Münzenberg M 2017 *J. Phys. D: Appl. Phys.* **50** 144003.

[8] Aqeel A, Mostovoy M, van Wees B J and Palstra T T M 2017 *J. Phys. D: Appl. Phys.* **50** 174006.

[9] Garst M, Waizner J and Grundler D 2017 *J. Phys. D: Appl. Phys.* **50** 293002.

[10] Sun C, Natterman T and Pokrovsky V L 2017 *J. Phys. D: Appl. Phys.* **50** 143002.

[11] Nakata K, Simon P and Loss D 2017 *J. Phys. D: Appl. Phys.* **50** 114004.